\documentclass[ 
onecolumn,
superscriptaddress, prl, tightenlines,
nofootinbib,
amsfonts,amsmath,amssymb]{revtex4}

\usepackage{bm} \usepackage{graphics,graphicx,epsfig,color}
\usepackage{subfigure} \usepackage{hyperref}

\begin{document}
 
\title{Atom gravimeters and gravitational redshift\footnote{Brief Communications Arising, Nature {\bf 467} (7311), E1, 2010; doi:10.1038/nature09340.}}

\author{Peter Wolf}\email{peter.wolf@obspm.fr} 
\affiliation{Syst\`emes de R\'ef\'erences Temps Espace, CNRS UMR 8630, Laboratoire National de m\'etrologie et d'Essais, Universit\'e Pierre et Marie Curie, Observatoire de Paris, 61 avenue de l'Observatoire, 75014 Paris, France.}
\author{Luc Blanchet}\email{blanchet@iap.fr}
\affiliation{GRECO, Institut d'Astrophysique de Paris, CNRS UMR 7095, UPMC, 98 bis boulevard Arago, 75014 Paris, France.}
\author{Christian J. Bord\'e}\email{chbo@ccr.jussieu.fr}
\affiliation{Syst\`emes de R\'ef\'erences Temps Espace, CNRS UMR 8630, Laboratoire National de m\'etrologie et d'Essais, Universit\'e Pierre et Marie Curie, Observatoire de Paris, 61 avenue de l'Observatoire, 75014 Paris, France.}
\author{Serge Reynaud}\email{serge.reynaud@upmc.fr}
\affiliation{Laboratoire Kastler Brossel, CNRS UMR 8552, \'Ecole Normale Sup\'erieure, \\UPMC, Campus Jussieu, 75252 Paris, France.}
\author{Christophe Salomon}\email{salomon@lkb.ens.fr}
\affiliation{Laboratoire Kastler Brossel et Coll\`ege de France, CNRS UMR 8552, ENS, UPMC, 24 rue Lhomond, 75231 Paris, France.}
\author{Claude Cohen-Tannoudji}\email{cct@lkb.ens.fr} 
\affiliation{Laboratoire Kastler Brossel et Coll\`ege de France, CNRS UMR 8552, ENS, UPMC, 24 rue Lhomond, 75231 Paris, France.}

\date{\today}

\begin{abstract}
In a recent paper, H. M\"uller, A. Peters and S. Chu [A precision measurement of the gravitational redshift by the interference of matter waves, Nature {\bf 463}, 926-929 (2010)] argued that atom interferometry experiments published a decade ago did in fact measure the gravitational redshift on the quantum clock operating at the very high Compton frequency associated with the rest mass of the C{\ae}sium atom. In the present Communication we show that this interpretation is incorrect.
\end{abstract}

\maketitle

In ref.~\cite{mueller} the authors present a re-interpretation of atom interferometry experiments published a decade ago~\cite{peters}. They now consider the atom interferometry experiments~\cite{peters} as a measurement of the gravitational redshift on the quantum clock operating at the Compton frequency $\omega_\text{C} = m c^2/\hbar \approx 2 \pi \times 3.0 \times 10^{25}\,\text{Hz}$, where $m$ is the C{\ae}sium (Cs) atom rest mass. They then argue that this redshift measurement compares favourably with existing~\cite{vessot} as well as projected~\cite{cacciapuoti} clock tests. Here we show that this interpretation is incorrect.

Since their publication, atom interferometry experiments of this type have been analysed as measuring the acceleration of free fall of Cs atoms in the interferometer, leading to a test of the universality of free fall (UFF) through a comparison with measurements using a freely falling corner cube. Although it remains less sensitive than tests using macroscopic test masses of different composition~\cite{williams,schlamminger} (relative precision $7 \times 10^{-9}$ versus $2 \times 10^{-13}$), this UFF test is very important because it is the most sensitive test that compares the free fall of quantum bodies with that of classical test masses.

But do such experiments measure the gravitational redshift? We first emphasize that the atom interferometer used in ref.~\cite{peters} is an accelerometer (or gravimeter). The associated signal is a phase shift $\Delta \varphi = \mathbf{k}\cdot\mathbf{g}\,T^2$, where $\mathbf{k}$ is the effective wavevector transferred by lasers, $\mathbf{g}$ is the local acceleration of gravity and $T$ is the interrogation time. It measures the acceleration of freely falling atoms, as defined with respect to the experimental platform that holds the optical and laser elements. With $\mathbf{k}$ and $T$ known from auxiliary measurements, one deduces the component of $\mathbf{g}$ along the direction of $\mathbf{k}$. If the whole instrument was put into a freely falling laboratory, the phase shift $\Delta \varphi$ would vanish.

The situation is completely different for instruments used for testing the universality of clock rates (UCR). An atomic clock delivers a periodic electromagnetic signal the frequency of which is actively controlled to remain tuned on an atomic transition. The clock frequency is sensitive to the gravitational potential $U$ and not to the local gravity field $\mathbf{g}=\bm{\nabla} U$. UCR tests are then performed by comparing clocks through the exchange of electromagnetic signals; if the clocks are at different gravitational potentials, this contributes to the relative frequency difference by $\Delta \nu/\nu=\Delta U/c^2$.

We show now that ref.~\cite{mueller} does not measure the gravitational redshift. This is clearly seen when evaluating the phase shift $\Delta \varphi$ using a Lagrangian formulation as employed in~\cite{mueller}; the Lagrangian may for example have the form considered in equation (8) of~\cite{mueller}. It can then be proved that the clock contribution $\Delta \varphi_\text{free}$ discussed in~\cite{mueller} --- that is, the difference of Compton phases along the two paths --- is exactly zero for a closed total path (equation (81) in ref.~\cite{storey}), or exactly cancelled by a second contribution associated with the splitting of the endpoints of these paths~\cite{borde}. The final phase shift $\Delta \varphi$ then arises entirely from a third contribution describing the interaction of light with atoms in the beam-splitting process. This explains why the phase shift takes the already discussed expression $\mathbf{k}\cdot\mathbf{g}\,T^2$, which does not depend on the Compton frequency $\omega_\text{C}$. 

The key point in this cancellation is a consistent calculation of the two paths in the atom interferometer as well as of the phases along these paths, both derived from the same Lagrangian by using the principle of least action in a standard way. When this is done in the case of~\cite{mueller}, we find that their equation (8) gives $\Delta \varphi_\text{free}=0$, irrespective of the value of the redshift violation parameter $\beta$. If one doubts the validity of this principle of least action, the significance and sensitivity of the test of ref.~\cite{mueller} remains to be evaluated~\cite{will}. 

To summarize, the experiment discussed in refs.~\cite{mueller} and \cite{peters} is an interesting test of UFF but not a measurement of the gravitational redshift or a test of UCR. Both kinds of tests (UFF and UCR) have their own merit, because they check in different ways whether all types of mass-energy are universally coupled to gravitation~\cite{damour}.

\end{document}